# Electronic phase separation at LaAlO$_3$/SrTiO$_3$ interfaces tunable by oxygen deficiency

V. N. Strocov,[1] A. Chikina,[1] M. Caputo,[1] M.-A. Husanu,[1,2] F. Bisti,[1] D. Bracher,[1] T. Schmitt,[1] F. Miletto Granozio,[3] C. A. F. Vaz,[1] and F. Lechermann[4]

[1]Swiss Light Source, Paul Scherrer Institute, CH-5232 Villigen-PSI, Switzerland

[2]National Institute of Materials Physics, Atomistilor 405A, RO-077125 Magurele, Romania

[3]CNR-SPIN, Complesso universitario di Monte S.Angelo, Via Cintia, I-80126 Naples, Italy

[4]Institut für Theoretische Physik, Universität Hamburg, Jungiusstr. 9, DE-20355 Hamburg, Germany

Electronic phase separation is crucial for the fascinating macroscopic properties of the LaAlO$_3$/SrTiO$_3$ (LAO/STO) paradigm oxide interface, including the coexistence of superconductivity and ferromagnetism. We investigate this phenomenon using angle-resolved photoelectron spectroscopy (ARPES) in the soft-X-ray energy range, where the enhanced probing depth combined with resonant photoexcitation allow access to fundamental electronic structure characteristics – momentum-resolved spectral function, dispersions and ordering of energy bands, Fermi surface – of buried interfaces. Our experiment uses X-ray irradiation of the LAO/STO interface to tune its oxygen deficiency, building up a dichotomic system where mobile weakly correlated Ti $t_{2g}$-electrons co-exist with localized strongly correlated Ti $e_g$-ones. The ARPES spectra dynamics under X-ray irradiation shows a gradual intensity increase under constant Luttinger count of the Fermi surface. This fact identifies electronic phase separation (EPS) where the mobile electrons accumulate in conducting puddles with fixed electronic structure embedded in an insulating host phase, and allows us to estimate the lateral fraction of these puddles. We discuss the physics of EPS invoking a theoretical picture of oxygen-vacancy clustering, promoted by the magnetism of the localized Ti $e_g$-electrons, and repelling of the mobile $t_{2g}$-electrons from these clusters. Our results on the irradiation-tuned EPS elucidate the intrinsic one taking place at the stoichiometric LAO/STO interfaces.

## Introduction

An inherent feature of many transition metal oxides (TMOs) is electronic phase separation (EPS), which is the spontaneous formation on the micro- to nano-scale of regions possessing different electronic and magnetic properties. This phenomenon is crucial, for example, for colossal magnetoresistance in manganites and the stripe order in cuprates [1–3]. The EPS phenomenon naturally propagates to interfaces of the TMOs. For the LAO/STO interface, the "drosophila" of oxide electronics where a mobile electron system (MES) is spontaneously formed – for entries see the reviews [4,5] – the EPS can be particularly strong because the huge dielectric constant of the nearly ferroelectric STO largely screens the electron repulsion and thus allows coagulation of large electron densities [6]. The EPS has been identified, for example, in a percolative metal-to-superconductor transition where a significant fraction of MES resists condensation to the superconducting ground state down to the lowest temperature [7,8]. On the magnetic side, ferromagnetic (FM) puddles embedded in metallic superconducting phase, observed in magnetotransport [9], magnetic torque magnetometry [10] and superconducting quantum interference device (SQUID) [11] experiments, explain the intriguing co-existence of the ferromagnetic and superconductive properties of the LAO/STO interface. Magnetoresistance, anomalous Hall effect and tunneling experiments [8,12,13] have also evidenced the presence of EPS at this interface. This phenomenon is extremely important from a device perspective [14] explaining why the interfacial charge carrier concentration $n_s$ observed in transport properties of the LAO/STO interface always falls short of predictions of the mean-field theories. Furthermore, the EPS can be tuned by an external electric field [6,15]. The first ARPES study of the EPS as co-existence of metallic and insulating phases has been performed by [16] on bare STO(100) and (111) surfaces as prototype of the LAO/STO interface.

Oxygen vacancies ($V_O$s) dramatically affect electronic and magnetic properties of all TMO systems. In a simplified picture of oxygen-deficient (OD) LAO/STO interfaces, each $V_O$ on the STO side releases two electrons. One fraction of them joins the MES formed by delocalized quasiparticles, which are Ti $t_{2g}$ derived, weakly correlated, non-magnetic and split by Rashba spin-orbit interaction [17], and form large polarons [18,19]. Sufficiently high concentrations of $V_O$s result in a a dimensionality transformation of the MES from 2D to quasi-3D [20–22]. Another fraction of electrons released by the $V_O$ stays near the Ti ion to form there a localized in-gap state (IGS) at binding energy $E_B \sim -1.3$ eV, which is Ti $e_g$ derived, strongly correlated, magnetic and is often viewed as a small polarons. These states determine the interfacial ferromagnetism of OD-LAO/STO [23,24]. This theoretical picture of the dichotomic electron system formed at the OD-LAO/STO interface [25–28] has recently been experimentally confirmed by resonant photoemission (ResPE) experiments [29]. The coexistence of the two radically different MES and IGS electron subsystems much enriches the physics of the OD-LAO/STO interfaces compared to the stoichiometric ones. As we will see below, the $V_O$s can also be used as a knob to tune the EPS, making the OD-LAO/STO interfaces the ideal platform to study this phenomenon.

Here, we demonstrate controllable tuning of the EPS at the OD-LAO/STO interfaces by generation of $V_O$s with intense soft X-rays from a synchrotron source. The electronic structure in this system is monitored by soft-X-ray ARPES (SX-ARPES), leveraging its enhanced probing depth and chemical specificity achieved by resonant photoexcitation. Following the evolution of the **k**-resolved electron dispersions and Fermi surface of the MES as a function of X-ray irradiation dose, we identify the EPS and estimate the lateral MES fraction developing with increase of the $V_O$-concentration. Finally, we elucidate a theoretical model of the EPS, which finds a clustering tendency of the $V_O$s and repulsion of the MES electrons from these clusters.

## Sample preparation and SX-ARPES experiment

OD-LAO/STO samples are a convenient platform to study the EPS. We have grown such samples on $TiO_2$-terminated STO(100) using Pulsed Laser Deposition (PLD). Before the LAO growth, the substrates were annealed in vacuum at 500°C for 30-40 min till the appearance of the characteristic RHEED pattern. Then a 4 u.c. thick layer of LAO was deposited at 800° C in $O_2$ pressure reduced to 8×10$^{-5}$ mbar. Whereas in the standard protocols [30] the growth is immediately followed by in-situ annealing in $O_2$ at 500°C, we cooled down the as-grown samples, and then post-annealed them ex-situ in $O_2$ at the same temperature for about 24 hours to ensure their complete saturation by oxygen [29].

Our **k**-resolved SX-ARPES experiments – for a recent review of this technique and its spectroscopic advantages see [31] – used resonant photoexcitation at the Ti *L*-edge that boosted photoemission response of the Ti-derived electron states at the buried interface. The measurements were performed at the SX-ARPES endstation [32] of the Advanced Resonant Spectroscopies (ADRESS) beamline [33] of the Swiss Light Source, Paul Scherrer Institute, Switzerland. The photon flux was around $10^{13}$ photons/s and focused into a spot of 30 x 75 µm$^2$ on the sample surface at an X-ray grazing incidence angle of 15°. The XAS data were measured in the total electron yield (TEY) mode. For the ARPES measurements, the combined (beamline and analyzer) energy resolution was set to ~50 meV. The sample temperature was kept at 12 K in order to suppress smearing of the coherent spectral structure [34] as well allow build-up of the $V_O$s. The series of Ti 2*p* core-level spectra depending on irradiation time ($t_{irr}$) were measured in intervals of 90 s, and ResPE spectra in intervals of 25 s. All series were averaged over 5 different points on the sample starting from fresh spots. Photoemission electron microscopy (PEEM) experiments aiming at spatial resolution of the nucleation and clustering of the $V_O$s (Supplemental 4) were carried out at the Surface and Interface Microscopy (SIM) beamline [35] of the Swiss Light Source.

$V_O$s on the STO side of the LAO/STO interfaces are readily identified by their characteristic signatures in the photoemission spectra, the IGS peak in the VB and $Ti^{3+}$ component in the Ti 2*p* core level spectra [19,29]. SX-ARPES characterization of our fresh samples did not detect these spectral structures. This evidences that the post-annealing quenched the $V_O$s in our samples below the sensitivity of our photoemission experiment of the order of 1%. It has been reported, however, that post-annealed samples deviate from the ST-ones on their transport properties [30]. We will see below that this deviation traces back not only to $n_s$ but also to nanoscale properties of the MES. Importantly, our samples appear susceptible to X-ray irradiation: Whereas the ST-samples are practically immune to it [18], ours, presumably due to non-perfect recovery of their crystallinity from the oxygen-deficient post-growth state, under X-ray irradiation were readily losing oxygen to form $V_O$s in STO, as evidenced by buildup of the IGS and $Ti^{3+}$ signals. Such dependence of the sample properties on the growth and annealing protocol goes along with the known sensitivity of STO substrates to their history with respect to the environmental conditions and temperature [36]. In particular, we have found that the rate of $V_O$-generation by X-rays strongly increases with the STO substrate temperature during its pre-growth annealing in vacuum, when $V_O$s might pre-form near its surface, and on $O_2$ pressure during the deposition of LAO. Interestingly, an increase of sample temperature above ~120K as well as X-ray irradiation of samples in $O_2$ pressure of about $10^{-7}$ mbar quench the X-ray generated $V_O$s. This can possibly be explained by the fact that X-ray irradiation cracks the physisorbed $O_2$ molecules into atomic oxygen [16,37] which can effectively penetrate into the LAO/STO heterostructure to quench the $V_O$s.

Secondary Ion Mass Spectrometry (SIMS) analysis of $O^{16}$ to $O^{18}$ isotope substitution under X-ray irradiation, reported elsewhere [22], suggests that the X-ray generated $V_O$s do not propagate beyond the top $TiO_2$-layer of STO. We note that the X-ray induced creation of the $V_O$ in STO implies efficient out-diffusion of oxygen through the LAO layer. Although well established down to room temperature [38], this process should however totally freeze at the sample temperature ~12K during our SX-ARPES experiments, where *kT* is 3 orders of magnitude smaller than the activation energy of $V_O$s. Such oddity is highly non-trivial and may suggests non-conventional oxygen migration mechanisms [22].

## *Overview of the electronic structure of OD-LAO/STO interfaces*

The electronic structure of the Ti-derived IGS- and MES-subsystems at the STO side of the OD-LAO/STO interface can be explored by ResPE at the Ti *L*-edge [19,39,40]. Fig. 1 (*a*) shows the experimental ResPE intensity as a function of $E_B$ and excitation energy (*hv*) measured after an X-ray irradiation for about 2 hours. Above the broad VB composed by the O 2*p* states in STO and LAO, the map shows the hallmark of the $V_O$s, the broad resonating peak at $E_B$ ~ -1.3 eV identifying the $e_g$-derived IGSs. The narrow resonating peak at $E_F$ identifies the $t_{2g}$-derived MES. For a detailed analysis of the ResPE behavior of OD-LAO/STO see [29].

The ARPES intensity images, taken at the $hv$-values marked at the above ResPE map, are shown in Fig. 1 (*b-d*). Measured with *s*-polarized incident X-rays, these images reflect the antisymmetric $d_{xy}$- and $d_{yz}$-bands sketched on top [18,41] as well as the disordered IGSs. The $d_{xy}$-band, due to its narrow interfacial localization, loses its photoexcitation cross-section at high energies [42] but is manifested by two bright points where it hybridizes with the $d_{yz}$-band due to the symmetry breaking caused by lattice distortions [18] as well as spin-orbit interaction [17]. The image (*b*) is taken at the Ti $L_3$-edge, where the $d_{xy}$ to $d_{yz}$ intensity ratio is higher compared to the $L_2$-edge [29], at $hv$ = 460.4 eV delivering sufficient separation of the MES peak from the Ti 2*p* core-level one excited by second-order X-ray radiation. The image (*c*) is characteristic of the broad dispersionless IGSs. It is taken at the $L_2$-edge (at the $L_3$-edge the IGSs overlap with the Ti 2*p* stray intensity) at $hv$ = 464.4 eV, where the IGS signal resonantly scales up and the MES one scales down. The image (*d*) measured at $hv$ = 466.4 eV is characteristic of the $d_{yz}$-band, whose intensity ratio to the $d_{xy}$-one X-ray enhances compared to the $L_2$-edge. To highlight the predominant spectral contribution, the images (*c-e*) and the corresponding $hv$ values on the ResPE map are marked as MES$_{xy}$, V$_O$ and MES$_{yz}$.

The band order and band dispersions observed in OD-LAO/STO adopt the same pattern as in ST-LAO/STO [18,41]. The band filling in OD-LAO/STO is somewhat larger, however, as characterized by the Fermi momentum $k_F$ of the $d_{yz}$ band in Fig. 1 (*d*) increasing to ~0.4 Å$^{-1}$ compared to ~0.33 Å$^{-1}$ for ST-LAO/STO [18,41]. Furthermore, the weight of the polaronic hump and thus renormalization of the effective mass $m^*$ reduce to ~1.5 compared to ~2.5 in ST-LAO/STO [18]. Importantly, although the V$_O$s generated by X-rays in STO are located predominantly in its top TiO$_2$-layer, analysis of out-of-plane ARPES dispersions [22] evidences that the resulting MES has actually quasi-3D character, with its penetrating depth into STO being more than 100 Å [21]. The large spatial extension of electron states generated by the localized V$_O$s is similar to the doping of conventional semiconductors with a tiny amount of foreign atoms.

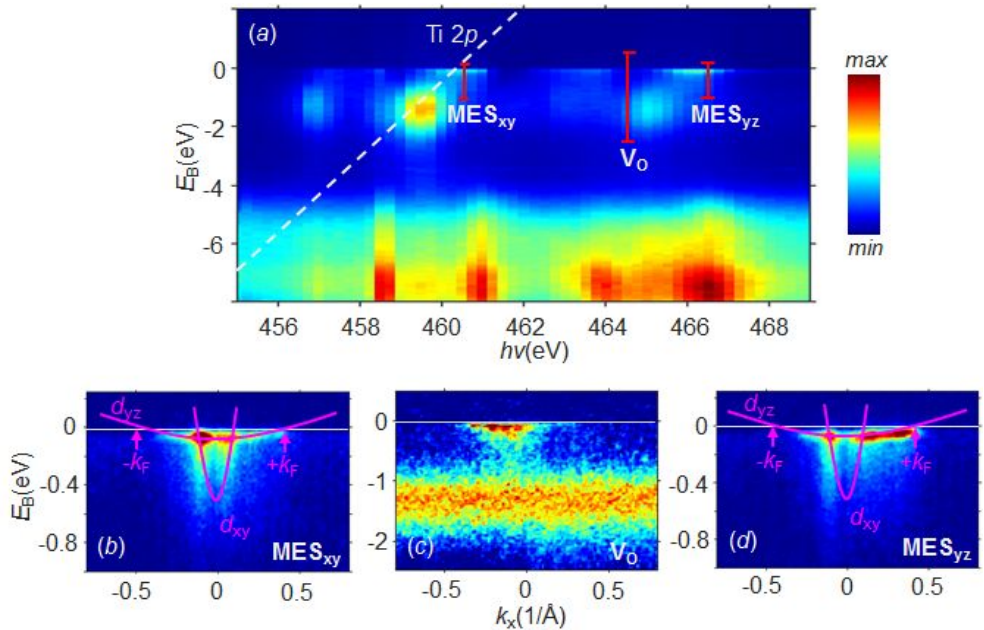

**Fig. 1**. ResPE of OD-LAO/STO: (*a*) angle-integrated ResPE intensity through the $L_3$ and $L_2$ edges. The IGS-peak at $E_B$~ -1.3 eV signals the V$_O$s; (*b-d*) ARPES intensity images measured at $hv$ = 460.4, 464.4 and 466.4 eV which enhance, respectively, the MES-$d_{xy}$ states, V$_O$-induced IGS, and MES-$d_{yz}$ states. These $hv$ values and corresponding $E_B$-intervals of the images are marked by red on (*a*). The $d_{xy}$- and $d_{yz}$-dispersions are sketched.

## Electronic phase separation

### X-ray irradiation dynamics

The effect of X-ray irradiation on the electronic structure of our OD-LAO/STO samples is illustrated in Fig. 2 by (downsampled) series of $t_{irr}$-dependent angle-integrated photoemission spectra collected in different spectral regions. Fig. 2 (*a*) shows the Ti 2*p* core-level spectra measured at $h\nu$ = 1000 eV. The increase of their $Ti^{3+}$ component manifests buildup of $V_O$s which reduce the $Ti^{4+}$ ions to the $3^+$ valency. The apparent $Ti^{3+}$ to $Ti^{4+}$ spectral weight ratio reaches ~0.1 at maximal $t_{irr}$. Taking into account the location of $V_O$s in predominantly in the top STO layer [22] and a photoelectron mean-free-path $\lambda$ of ~11 Å at our kinetic energy [37,43], this figure indicates that about 30% of the Ti ions in this layer become $Ti^{3+}$. The panels (*b-c*) shows $t_{irr}$-dependent angle-integrated ResPE spectra of the VB integrated through the first Brillouin zone (BZ). The series in (*b*) was measured at $h\nu$ = 464.4 eV which boosts the IGS-peak at $E_B$ ~ -1.3 eV, see Fig. 1 (*a*). Negligible at the zero-$t_{irr}$ limit, its intensity and thus IGS-concentration gradually increases under irradiation. The series (*c*) and (*d*) show development of the MES signal at $h\nu$ = 460.4 and 466.4 eV marked as $MES_{xy}$ and $MES_{yz}$, respectively, to highlight the predominant spectral contribution. The increase of this signal with $t_{irr}$ manifests the increase of the integral MES concentration, following the $V_O$ buildup. The observed evolution of both IGS and MES peaks reflects the fact that the $V_O$s inject localized electrons into the IGSs and mobile ones into the MES. As a side effect of the $V_O$ accumulation, we observe that the polaronic tail of the MES quasiparticle peak [18] notably reduces its spectral weight under irradiation due to increase of the MES concentration that screens the electron-phonon interaction.

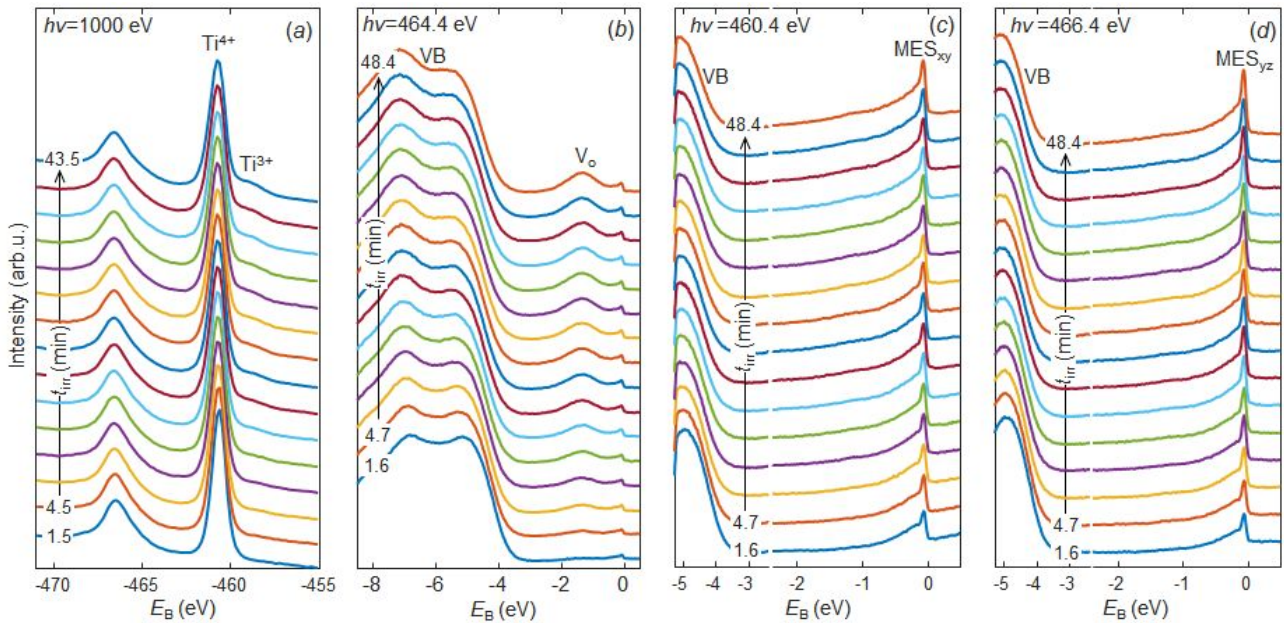

**Fig. 2.** X-ray irradiation effect on electronic structure of the OD-LAO/STO interface: (*a*) Ti 2*p* core-level spectra, where the $Ti^{3+}$ component corresponds to the $V_O$s; ResPE spectra measured (*b*) through the whole VB at $h\nu$ = 464.4 eV boosting the IGS-peak, and (*c-d*) at $h\nu$ = 460.4 and 466.4 eV boosting response of the $d_{xy}$ and $d_{yz}$ states, respectively. Note different energy scale of the VB vs $V_O$ and MES regions in (*c-d*). The experimental series are downsampled to the indicated time intervals. Under X-ray irradiation the $Ti^{3+}$ core level, IGS- and MES-intensity increase, following buildup of the $V_O$s.

The $Ti^{3+}$, IGS and $MES_{xy}/MES_{yz}$ spectral weights, evaluated from the data in Fig. 2, are presented in Fig. 3 (*a*) as a function of $t_{irr}$. To separate the IGS spectral contribution from the overlapping MES, we have averaged the ARPES intensity over the angular intervals outside the MES. The difference of these out-of-MES spectra from the zero-$t_{irr}$ one, where the IGS-signal was negligible, identified the IGS-spectra for every $t_{irr}$. The IGS-weight $W_{IGS}$ was then obtained by integration over the energy width of the IGS-peak. To separate the MES contribution, in turn, we have averaged the ARPES intensity over the angular interval inside the MES and subtracted the corresponding out-of-MES spectra. The MES weight $W_{MES}$ was determined by integration over the whole MES bandwidth including the polaronic tail. The two MES curves in Fig. 3 (*a*) correspond to the two $h\nu$ values which boost the contribution of either $d_{xy}$ or $d_{yz}$ states in the total ARPES response. All curves are normalized here to their (statistically significant) value at maximal $t_{irr}$.

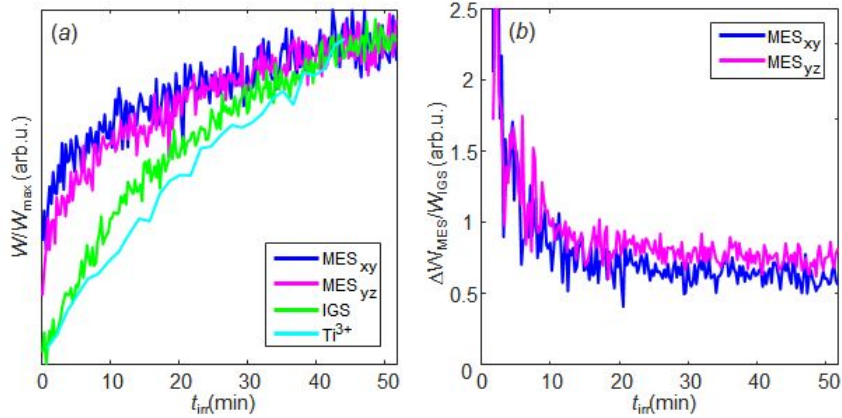

**Fig. 3.** (*a*) $Ti^{3+}$, IGS and MES spectral weights, normalized to their maximal value, as a function of irradiation time; (*b*) The MES-to-IGS buildup ratio. The faster saturation of the MES-signal compared to the IGS-one indicates that at large $V_O$ concentration the doped electrons preferentially populate the IGSs.

The $t_{irr}$-dependencies in Fig. 3 (*a*) bear two important facts: (1) In the zero-$t_{irr}$ limit, $W_{MES}$ at both resonances extrapolates to certain non-zero value. In other words, the LAO/STO interface is conductive already without the $V_O$s. This fact supports the electrostatic scenarios of the LAO/STO conductivity as resulting from a redistribution of electrons to the interface to minimize the electrostatic energy created by polar discontinuity between LAO and STO. In these scenarios the MES electrons are injected by either states from the VB maximum of LAO pushed above $E_F$ – the well-known "polar catastrophe" – or from localized defect states in LAO created, for example, again by $V_O$s which existed there before irradiation [21,37,44]; (2) As we have already mentioned, in a simplistic picture, one of the two electrons doped by each $V_O$ localizes at the IGS, and another joins the mobile MES. Fig. 3 (*a*) shows that whereas after ~10 min of irradiation the production of both IGS- and MES-electrons slows down towards saturation, for the MES the slowdown is stronger. This trend is emphasized in Fig. 3 (*b*) as a ratio of the $W_{MES}$ buildup $\Delta W_{MES}$ (relative to the zero-$t_{irr}$ weight $I_0$) to $W_{IGS}$. This ratio dramatically drops towards $t_{irr}$~10 min, indicating that at large $V_O$-concentration the doped electrons tend to localize in the IGSs rather than escape into the MES. This observation still has to be fully understood, although a few theoretical works have demonstrated that properties of the $V_O$s depend on their concentration regime [21,45]. The first ARPES study of the EPS at the bare STO(100) and (111) surfaces [16,28] has presented evidence that this phenomenon is driven by clustering of VOs. Indeed, LDA+*U* calculations on bare STO surfaces [21,45] suggest that the electron distribution between the MES and IGSs may depend on particular configurations of $V_O$s; while isolated $V_O$s on the surface tend to donate both released electrons into the MES, their cluster configurations favour even distribution.

*Identification of EPS*

We will now turn to the X-ray irradiation dynamics of **k**-resolved electron dispersions $E(\mathbf{k})$ and Fermi surface (FS). The progressive increase of the MES spectral intensity with irradiation, seen in Fig. 2 (*c*,*d*) and quantified by $W_{MES}$ in Fig. 3 (*a*), evidences increase of the total concentration $n_s^{tot}$ of the MES electrons. The first idea might be that this is caused by a progressive increase of the band population. The experimental **k**-resolved ARPES images acquired at small and large $t_{irr}$ are displayed in Fig. 4 (*a*) and (*b*). We observe, surprisingly, that these images demonstrate identical dispersions and population of the light $d_{xy}$ and heavy $d_{yz}$-bands, with the only difference being a rigid increase of their spectral intensity (slight differences in the polaronic tale are beyond the scope of this work). The same behavior can be seen in ARPES data taken in other **k**-space regions (see Supplemental 2).

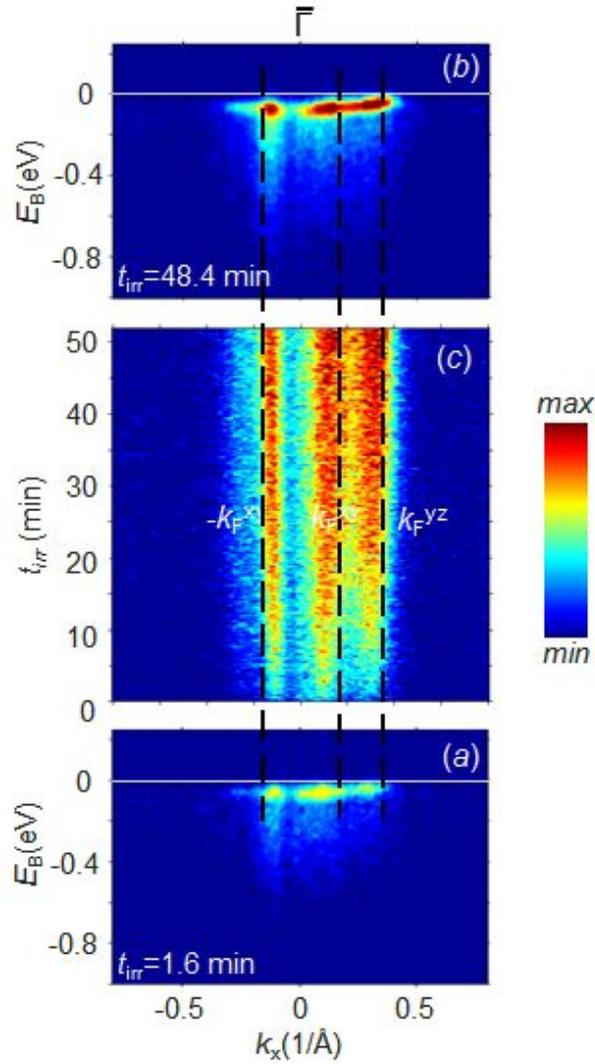

**Fig. 4.** Irradiation dependence of **k**-resolved electronic structure: (*a,b*) ARPES images of *E*(**k**) at small and large $t_{irr}$ (indicated). (*c*) Fermi intensity as a function of $t_{irr}$. The $k_F$ values corresponding to the $d_{xy}$- and $d_{yz}$-bands are marked through with dashed lines ($-k_F^{yz}$ is not visible because of the vanishing matrix element). The spectral intensity increase at constant band structure and $k_F$ manifests the EPS.

Furthermore, we have quantified $k_F$ of the experimental bands as corresponding to maxima of the $-dI_F/d|k|$ negative gradient of the Fermi intensity. This method [46] is particularly suited for our case, where the bandwidth is comparable with the energy resolution, and the spectral intensity is significantly distorted by matrix-element variations. We find, remarkably, that the small- and large-$t_{irr}$ ARPES images in Fig. 4 (*a,b*) show identical $k_F$ ~ 0.4 Å$^{-1}$ (marked by dashed lines) even for the flat $d_{yz}$-band most sensitive to its population. This fact identifies the same Luttinger counts $n_s^{Lutt}$ of the FS embedding the MES electrons. Extending this observation, Fig. 4 (*c*) shows a map of $I_F$ through the whole range of $t_{irr}$. The $k_F$ values determined by the gradient method (dashed lines, for the corresponding $-dI_F/d|k|$ map see Supplemental 2) are again constant, identifying constant $n_s^{Lutt}$ for both $d_{xy}$ and $d_{yz}$ bands. The statistics of our experimental $I_F$ and $-dI_F/d|k|$ maps shows this trend undoubtedly through the whole range of $t_{irr}$ except perhaps $k_F$ of the $d_{yz}$-band in the region of very small $t_{irr}$ where the ARPES signal is yet small.

The observed discrepancy between the total $n_s^{tot}$, increasing with $t_{irr}$, and constant $n_s^{Lutt}$ can be reconciled within the EPS picture, where conducting puddles at the interface co-exist with insulating ones. The MES signal in the ARPES spectra comes only from the conductive puddles, while the non-conductive ones stay silent. Upon X-ray irradiation and concomitant increase of the $V_O$-concentration, the conducting puddles inflate their total interfacial fraction, as evidenced by scaling up of $n_s^{tot}$, whereas their local electronic structure stays the same as evidenced by their constant band structure and $n_s^{Lutt}$ seen in ARPES. Due to the EPS, $n_s^{Lutt}$ always promises larger $n_s^{tot}$ than will actually be seen in electron transport. A sketch of the EPS in OD-LAO/STO, where the MES puddles have a quasi-3D character, is shown in Fig. 5.

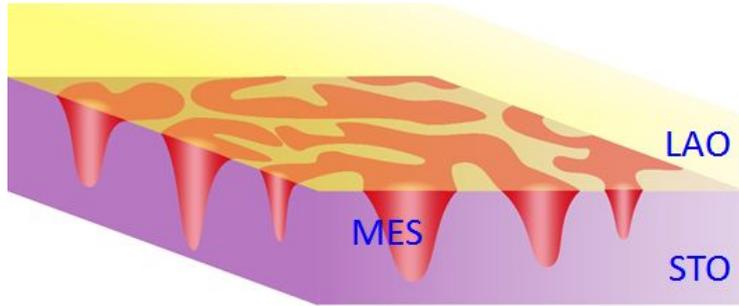

**Fig. 5.** Schematic representation of quasi-3D conducting puddles of MES in OD-LAO/STO.

The depth extension of MES in the OD-LAO/STO samples [20,21,47] is much more than $\lambda$ ~11 Å at our kinetic energy [37,43]. Therefore, the observed ARPES intensity is roughly proportional to the lateral fraction of the conducting puddles, which we will call the lateral conducting fraction (LCF). In our case, the irradiation dynamics of the MES signal in Fig. 3 (*a*) shows that from small to large $t_{irr}$ the LCF increases by a factor ~3.3. As the LCF by definition can not exceed 100%, we obtain its upper limit back to small oxygen deficiency as ~30%. An alternative way to assess the EPS, which is based on the MES intensity comparison to the Ti 2*p* peak, yields a similar estimate of the LCF (see Supplemental 3).

We note that the observed constant band structure parameters of the MES suggest that it has the same quasi-3D character already in the fresh samples. Indeed, the detection limit of our SX-ARPES experiment of ~1% corresponds to a spatial separation of the $V_O$s of ~40 Å, which is well within the extension of the MES wavefunctions of more than 100 Å [21]. Therefore, the spatial extension and thus dimensionality of the MES settle already at minute concentrations of $V_O$s.

## Clustering of $V_O$s and its role in EPS

The parallel development of the IGS-and MES-signals with irradiation, observed in our experiments, suggests that the $V_O$s play an important role in the EPS. Indeed, theoretical analysis of interplay between the $t_{2g}$ states of the MES with the orbitally reconstructed $e_g$ states formed by the $V_O$s yields a rich phase diagram that includes regions of phase-separated magnetic states [25]. Furthermore, previous theoretical works predicted a tendency of $V_O$s to form clusters [48,49] which should unavoidably affect the EPS.

Our experimental data cannot directly elucidate the $V_O$-clustering tendency at the OD-LAO/STO interfaces. However, we can approach it via basic arguments from the alloy theory. Its cluster-expansion formalism provides a versatile tool to map lattice-configurational energies onto effective (Ising-like) interactions also for vacancy problems [50,51]. Here, we have employed density functional theory (DFT) as well as DFT plus dynamical-mean field theory (DFT+DMFT) – see [29] for particular computational details – to determine the respective configurational energies for an LAO/STO supercell. Based on these calculations, we have performed the binary-alloy mapping in the simplest nearest-neighbor pair approximation within an interface square lattice of oxygen ions (Fig. 6) where the site A was set to the O atom and B to the $V_O$. This picture yields an effective interaction integral $J_{eff} = ¼(E_{AA}-2E_{AB}+E_{BB})$, where $E_{AA}$ stands for the stoichiometric supercell energy, $E_{AB}$ for the supercell with one $V_O$, and $E_{BB}$ for the one with a nearest-neighbor double $V_O$s. Just as in the Ising theory of ordered magnetism, a positive $J_{eff}$ drives alloy ordering ("antiferromagnetism"), while a negative one drives phase separation ("ferromagnetism"). Indeed, our calculations have revealed that in the LAO/STO case the effective pair interaction turns out negative, i.e. $J_{eff}$=-143meV in the DFT framework and -155meV in DFT+DMFT. Although a more thorough inclusion of further long-range interactions is needed for an accurate description of the defect interactions, this simple approximation already provides a good clue for a basic tendency towards clustering of $V_O$s in OD-LAO/STO. Our theoretical picture supports the conjecture of Li et al. [10] that the inhomogeneous distribution of $V_O$s stabilizes the magnetic order at the interface. On the experimental side, our irradiation-dependent photoemission electron microscopy (PEEM) measurements have shown that the size of the $V_O$-clusters does not exceed 500 Å (see Supplemental 4).

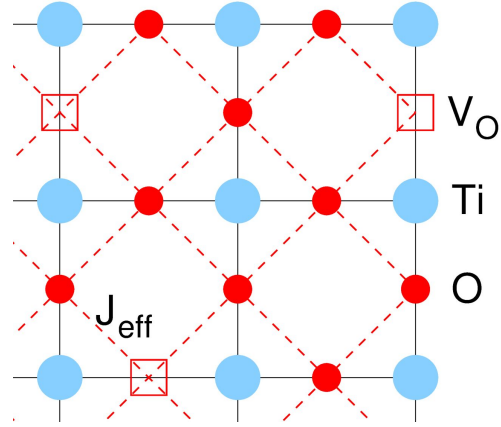

**Fig. 6.** Interface square lattice of oxygen ions in LAO/STO used in the $J_{eff}$ analysis. A negative $J_{eff}$ value favours clustering of $V_O$s.

The next important question on the physics of EPS that our experimental data cannot directly address is how are the MES puddles located relative to the $V_O$-clusters. One can however approach this question theoretically by considering which Ti 3$d$ orbitals are involved in the MES vs $V_O$-regions. Once the $V_O$-clustering sets in, the orbitals mainly occupied in the $V_O$-regions are the strongly localized Ti $e_g$-ones, whereas in the fairly defect-free MES-regions these orbitals are empty and only the strongly itinerant Ti $t_{2g}$ orbitals are slightly occupied. Therefore, the MES tends to spatially separate from the VO-clusters. Surely, as it was shown recently [29], there is still a coupling between both orbital degrees of freedom, yet in main terms, the spatial separation of the MES from the $V_O$s via the distinct orbital characteristics carries over the $V_O$-clustering into the electronic domain.

The formation of $V_O$-clusters is also important for the magnetism of LAO/STO. Recent computational analysis including many-body correlations [23,24] suggest that already single-$V_O$ configurations can build up the FM order. As observed in a theoretical study of the oxygen-deficient STO surface [27], the $V_O$ clustering can promote, furthermore, a FM-arrangement of the local moments originating from the nearby Ti sites.

*EPS in ST-LAO/STO and discussion*

How does the irradiation-tuned EPS observed at the OD-LAO/STO interfaces relate to the ST-ones? Because of the different MES-dimensionality, the present results can not be directly extrapolated to ST-LAO/STO. In that case, however, we can approach the intrinsic EPS by comparing the FS seen in the ARPES experiment with transport data. The previous SX-ARPES studies on ST-LAO/STO [18,19,41] find the FS where, evaluated from the d$I_F$/d$|k|$ minima, $k_F$ of the $d_{xy}$-derived circular sheet is ~0.07 Å$^{-1}$, and those of the $d_{xz/yz}$-derived elliptical sheets are ~0.33 Å$^{-1}$ and ~0.12 Å$^{-1}$ along the long and short axes, respectively. These values yield the Luttinger count $n_s^{Lutt}$ ~ 7·10$^{13}$ e/cm$^2$ which, because of the neglect of the smaller FS sheets derived from the $d_{xy}$ states in deeper STO layers, should be considered as the lower estimate. The difference of this figure from $n_s$ between 4 and 6·10$^{13}$ detected by Hall measurements [41] identifies an EPS with a significant LCF of the order of 70%. This identification is corroborated by previous DFT calculations fitted to the transport $n_s$ which, missing the EPS, produced systematically lower band energies (to embed more electrons) compared to the SX-ARPES experiment [41].

Similar reduction of $n_s$ seen in the Hall effect has been noticed in comparison with results of inline electron holography [52]. We note that the experimental Luttinger count, characterizing the MES regions, is by a factor of almost 5 smaller than $n_s$ = 3.3x10$^{14}$ $e$/cm$^2$ corresponding to 0.5 $e$/u.c. of the ideal "polar catastrophe" scenario.

Our spectroscopic results on the EPS at the OD-LAO/STO interfaces and their above extension to the ST-ones are consistent with a number of previous reports. For bare STO surfaces, the EPS developing under X-ray irradiation has been identified in an angle-integrated photoemission study [16,28]. For the LAO/STO interfaces, our area-sensitive spectroscopic evidence of the EPS is complemented by local methods such as scanning tunneling microscopy/spectroscopy [53,54] and atomic force microscopy [15] as well as from transport and magnetic properties [6,9]. Furthermore, our experimental results demonstrate that not only the electron dispersions and band population but also the energy broadening of the spectral peaks are independent of the interfacial conducting fraction up to its saturation. This behavior suggests that the interface separation into the conducting and non-conducting areas is sharp, without any significant variation of electronic structure between different MES puddles. This fact casts doubt on previous conjectures that the EPS could be associated with variation of the band population along the interface, see [9], for example. On the theoretical side, Scopigno et al. [6] suggested that the EPS in ST-LAO/STO results from a lateral confinement tendency of the MES to avoid a thermodynamically unstable state with negative compressibility. Other non-$V_O$ scenarios invoked formation of a Jahn-Teller polaronic phase [55], Rashba spin-orbit coupling [56,57] and even superconducting pairing interaction [57].

## *Summary and outlook*

In summary, we have demonstrated EPS at LAO/STO interfaces tunable through their oxygen deficiency. Our experiment used ResPES at the Ti 2$p$ edge to follow the evolution of the interfacial electronic structure as a function of $V_O$-concentration gradually developing under X-ray irradiation. The $V_O$s generated by X-ray build up a dichotomic electron system where weakly correlated and non-magnetic delocalized MES-electrons coexist with strongly correlated and magnetic localized IGS-electrons. The observed irradiation dynamics of the OD-LAO/STO electronic structure identifies the EPS as accumulation of the MES-electrons in quasi-3D conducting puddles with fixed electronic structure, which are embedded in the insulating phase. Our theoretical analysis shows that the $V_O$s, promoted by the magnetism of the IGS-electrons, tend to form clusters, and the MES puddles tend to spatially separate from them. Our PEEM data sets the upper limit on the $V_O$-cluster size as <500 Å. The intricate physics of the EPS, involving the interplay of the MES and IGSs, require further theoretical and experimental efforts where various area-sensitive spectroscopies such as ARPES are combined with local probes such as scanning tunneling microscopy/spectroscopy. Tunability of the EPS through oxygen deficiency can be used, on one end, for manipulation of electron transport properties of forthcoming oxide-based devices and, on the other end, their magnetic functionality. In this case the device heterostructures can be patterned using X-ray or e-beam lithography. The EPS persists intrinsically in the paradigm ST-LAO/STO case, with the LCF of the order of 70%.

The pronounced polaronic effects and the EPS are two native demerits of STO-based systems which fundamentally restrict, respectively, the mobility and concentration of charge carriers even under optimal tuning with oxygen deficiency. This fact calls for a search for other materials to replace STO as a standard platform for oxide electronics. The EPS can however be turned into an advantage of STO-based materials, where their natural separation into superconducting and ferromagnetic areas might allow an intrinsic realization of new device functionalities such as networks of ferromagnetic Josephson junctions.

## **Acknowledgments**

We thank R. Claessen, C. Cancellieri, M. Radovic, N. Pryds, U. Aschauer, G. Drera and R. De Souza for sharing fruitful discussions. F.L. acknowledges financial support from the DFG project LE 2446/4-1. DFT+DMFT computations were performed at the JUWELS Cluster of the Juelich Supercomputing Centre (JSC) under project number hhh08. A.C. and M.C. acknowledge funding from the Swiss National Science Foundation under grant no. 200021_165529.

# Supplemental Information

## 1. X-ray induced energy shifts of the ARPES spectral structures

The intensity evolution of the ARPES spectral structures under X-ray irradiation is accompanied by their energy shifts. Fig. S1 (*c*) compiles the $E_B$-shifts of the Ti$^{4+}$ core-level peak, two VB-ones and IGS-one from their position at saturation, with the MES-peak pinned at $E_F$. A small Ti$^{4+}$ downward shift of ~80 meV shows deepening of electrostatic potential in the interface region with irradiation. Both VB peaks shift down by ~350 meV, which much exceeds an energy shift of the Ti-derived VB states in STO expected from the Ti$^{4+}$ irradiation dependence. As suggested by Gabel et al. (Phys. Rev. B **95** (2017) 195109), this effect may result from the VB contribution coming from the LAO overlayer whose band alignment with STO depends on the developing MES concentration on the STO side. The IGS energy position, interestingly, does not show any clear evolution within the experimental statistics and appears to stay at fixed energy position relative to the MES at $E_F$.

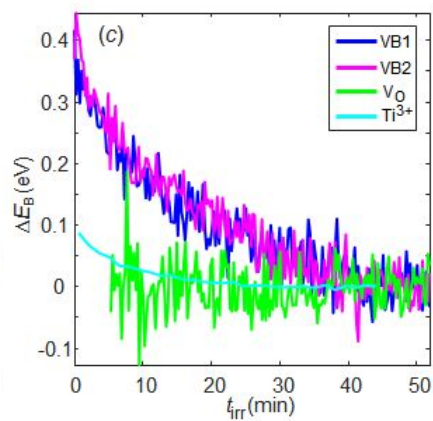

**Fig. S1.** Energy shifts of the Ti$^{4+}$ and VB spectral peaks from Fig. 2 (*a*,*b*) with irradiation.

## 2. X-ray irradiation dynamics of band structure and Fermi surface in an extended k-space region

Fig. S2 (*a,b*) represent ARPES images of our OD-LAO/STO samples at small and large $t_{irr}$, respectively, measured across two BZs. Similarly to Fig. 4, these images show $E(\mathbf{k})$ independent of $t_{irr}$. (*c,d*) show the corresponding Fermi intensity $I_F$ and its negative gradient $-dI_F/d|k|$. The maxima of the latter, going along vertical lines, show constant $k_F$ and thus constant local $n_s$. This characteristic pattern identifies the EPS.

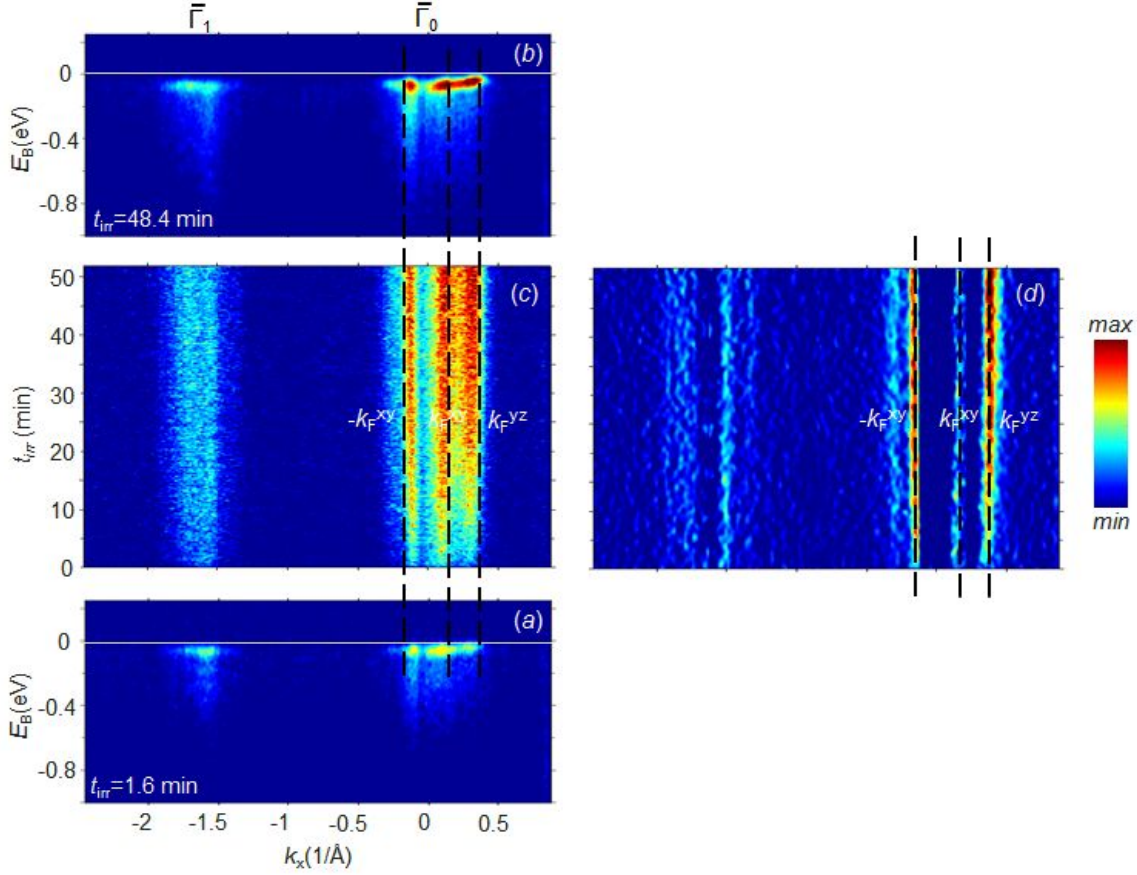

**Fig. S2**. Irradiation dependence of electronic structure through two BZs: (*a,b*) ARPES images of $E(\mathbf{k})$ at small and large $t_{irr}$, respectively; (*c,d*) Fermi intensity $I_F$ and its negative gradient $-dI_F/d|k|$, respectively, as a function of $t_{irr}$. $k_F$ of the $d_{xy}$- and $d_{yz}$-bands determined from the $-dI_F/d|k|$ maxima are marked through with red dashed lines.

## 3. Estimate of the LCF based on intensity ratio of the MES and Ti 2p signals

A way to estimate the irradiation-dependent LCF in the OD-LAO/STO samples can be based on comparison of the MES spectral intensity to the Ti 2p peak as the intensity reference. In our ResPE map, Fig. 1 (*a*), the Ti 2p peaks excited by second-order light (photon energy 2*hv*) from the beamline monochromator go along the ascending dashed line. Fig. S3 presents the $E_F$-region, including the second-order Ti 2p peak, of the angle-integrated spectra measured at *hv*=460.4 eV, corresponding to Fig. 3 (*c*), at small and large $t_{irr}$. The spectra are angle-integrated within two BZs, normalized to the Ti 2p peak intensity, and offset for visual clarity. The small-$t_{irr}$ spectrum shows the MES to Ti 2p integral intensity ratio smaller by a factor of ~2.5 compared to the large-$t_{irr}$ which can be considered as a reference of the LCF ~ 100%. This ratio yields the upper limit of the LSF as about 40%, in good agreement with the above analysis based on the MES dynamics only.

Also shown in Fig. S3 is the corresponding spectrum for an ST-sample measured at *hv*=460.4 eV (C. Cancellieri et al. Nature Comm. **7** (2016) 10386). The first interesting observation is that the weight of the polaronic hump (marked LO3) for the ST- and small-$t_{irr}$ OD-sample is similar, and at large $t_{irr}$ the polaronic weight significantly reduces. The ST-sample shows the MES to Ti 2p intensity ratio smaller by a factor of ~2.2 compared to the large-$t_{irr}$ OD-one. We also recall that the OD-samples show an apparently larger FS in comparison with the ST-ones. However, these observations can not directly lead to the LCF estimate for the ST-samples because of the different MES dimensionality for the ST- and OD-LAO/STO.

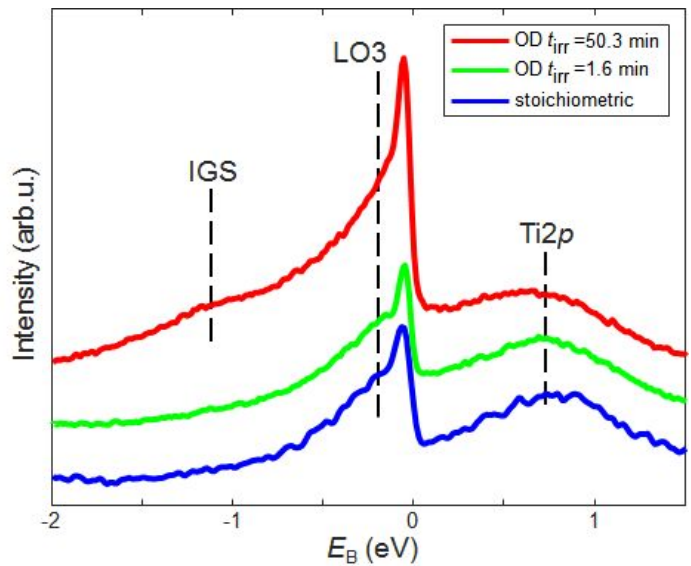

**Fig. S3.** Angle-integrated spectra at *hv*=460.4 eV in the region near $E_F$ including the second-order Ti 2p peak for the ST- and OD-LAO/STO samples. The MES to Ti 2p integral intensity ratio is related to the LCF.

## 4. Microscopic investigation on the EPS at the LAO/STO interface

Our X-ray irradiation dependent PEEM measurements aimed spatial resolution of the nucleation and clustering of the $V_O$s as marked by the $Ti^{3+}$ ions. The experiment was carried out at the SIM beamline of the Swiss Light Source which delivered a photon flux of ~$10^{14}$ ph/sec/0.01% within a spot of 100x200 µm². The measurements were performed on two LAO/STO samples, prepared as described above, which were cooled to 110K. In few different spots over these samples, the following data series have been acquired: a XAS spectrum through the Ti $L_{3-2}$ edge, a sequence of X-ray irradiation dependent PEEM images of secondary electron emission at two specific photon energies, and the concluding XAS spectrum. The two photon energies used to collect the images were $hv$ = 458.9 and 459.8 eV, corresponding to the Ti $2p_{3/2}$ -> Ti $3d$ transitions of, respectively, the $Ti^{4+}$ and $Ti^{3+}$ ions.

Fig. S4 shows the typical behavior found in all experimental series: the differential images of the $Ti^{4+}$ and $Ti^{3+}$ specific PEEM snapshots acquired at different $t_{irr}$ (a-c) did not show any sign of formation of islands of different Ti chemical state. However, comparing the two XAS spectra performed at the beginning and the end of the image sequence (d) shows that a sizable $Ti^{3+}$ component develops under the beam. Furthermore, the ratio of the $Ti^{3+}$ and $Ti^{4+}$ signals integrated through the whole PEEM detector area (e) linearly increases with $t_{irr}$. These findings demonstrate that the dimensions of the $Ti^{3+}$ puddles stay at least less than the instrumental lateral resolution of 500 Å.

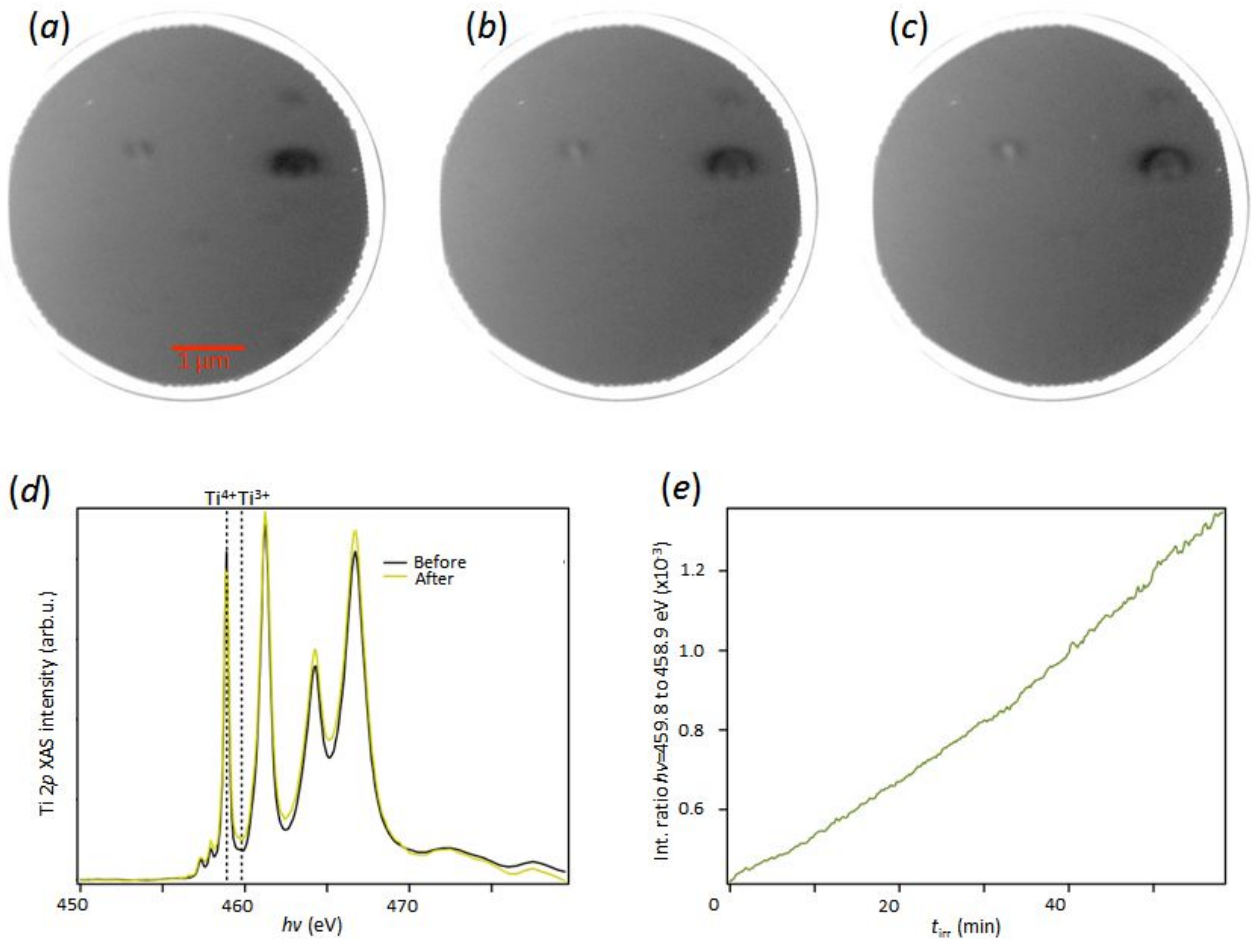

**Fig. S4.** PEEM measurements on OD-LAO/STO samples: (a-c) Differential images of the $Ti^{4+}$ and $Ti^{3+}$ specific PEEM snapshots acquired at different $t_{irr}$; (d) XAS spectra before and after acquiring the $t_{irr}$-dependence, were the two $hv$-values specific of $Ti^{3+}$ to $Ti^{4+}$ marked by dashed lines.